\documentclass[10pt, twocolumn]{article}
\usepackage{style/latex8}
\usepackage{times}
\usepackage{multirow}
\usepackage{url}
\usepackage[open-square,italic-theorems]{QED}

\usepackage[normalem]{ulem}

\usepackage{color}

\usepackage{fancyhdr}
\usepackage{mathptmx}
\usepackage{latexsym}
\usepackage{amssymb}
\usepackage{times}
\usepackage{epsfig}
\usepackage{subfigure}
\usepackage{graphicx}
\usepackage{algorithmicx}
\usepackage[ruled]{algorithm}
\usepackage{algorithm}
\usepackage{algpseudocode}

\renewcommand{\algorithmiccomment}[1]{{{\small \em// #1}}}

\usepackage{glossaries}

\newacronym{ris}{RIS}{Reconfigurable Intelligent Surface}
\newacronym{edhoc}{EDHOC}{Ephemeral Diffie-Hellman Over COSE}
\newacronym{pcs}{PCS}{Post-Compromise Security}
\newacronym{li}{LI}{Lawful Interception}
\newacronym{aka}{AKA}{Authentication and Key Agreement}
\newacronym{rl}{RL}{Reinforcement Learning}
\newacronym{los}{LoS}{Line Of Sight}
\newacronym{psk}{PSK}{Pre-Shared Key}
\newacronym{pki}{PKI}{Public Key Infrastructure}

\begin{document}

\title{\vspace{.75cm}Man-in-the-Middle Proof-of-Concept via
Krontiris'\\ Ephemeral Diffie-Hellman Over COSE (EDHOC) in C\thanks{This is a selected and revised version of Chapter 3 from Daniel Hennig's internship 
manuscript on `Cybersecurity Impact of Malicious Reconfigurable Intelligent 
Surfaces’, reproducing part of the work carried out during a five-month research 
internship (from April 2025 to September 2025) at the SAMOVAR laboratory 
of Télécom SudParis, within the activities of the SCN team.}}

\author{
  \vspace{.45cm}
  Daniel Hennig$^{\dagger,\ddagger}$ and Joaquin Garcia-Alfaro$^{\dagger}$\\
  $^{\dagger}$SAMOVAR, Télécom SudParis, Institut Polytechnique de Paris, Palaiseau, France\\
  $^\ddagger$INSA Toulouse, Département de génie electrique et informatique, France \\
  E-mail: {\it hennig@insa-toulouse.fr,}\\
  {\it joaquin.garcia\_alfaro@telecom-sudparis.eu}
}

\pagestyle{fancy}
\lfoot{Technical Report, Télécom SudParis}
\cfoot{\thepage}
\rfoot{2024-2025}
\fancyhead{}
\renewcommand{\headrulewidth}{0pt}

\maketitle

\thispagestyle{fancy}

\begin{abstract}
\noindent This report presents part of the work carried out during a five-month research internship at the SAMOVAR laboratory of Télécom SudParis, focusing on some security aspects of B5G (Beyond 5G) networks. The internship combined literature review, protocol 
analysis, and simulation work. Particular attention was given to the authentication process of the \gls{edhoc} lightweight key exchange protocol, examining how Man-in-the-Middle (MitM) attacks could undermine trust models, e.g., under the scope of lawful interception and its risk to facilitate mass surveillance. We report only some technical aspects associated to the internship under the tasks associated to the aforementioned MitM attack scenario designed and implemented against the \gls{edhoc} protocol. Some other specific aspects of the work, mainly focusing on the security implications of malicious metasurfaces against B5G networks, are excluded from the scope of this report.
\end{abstract}

\section{Introduction}
\label{sec:intro}

\begin{table*}[!t]
	\caption{\gls{edhoc} authentication methods}
	\label{tab:edhocauth}
	\centering
	\begin{tabular}{|l|c|c|c|c|}
			\hline
			     \textbf{Method type value} & \textbf{Initiator Authentication Key} & \textbf{Responder Authentication Key}\\
			\hline
			0 & Signature Key & Signature Key \\
			\hline
			1 & Signature Key & Static DH Key \\
			\hline
			2 & Static DH Key & Signature Key \\
			\hline
		      3 & Static DH Key & Static DH Key \\
            \hline
            4 & Pre-Shared Key & Pre-Shared Key\\
            \hline  
        \end{tabular}
\end{table*}

\noindent The Ephemeral Diffie-Hellman Over COSE (\gls{edhoc}) protocol is a lightweight key exchange protocol for constrained devices (e.g., network nodes running on limited amount of memory space) or for scenarios under constrained network properties (e.g., in terms of bandwidth). It achieves the aforementioned goal by reducing the amount of messages sent, the message size, as well as by reusing primitives already used by other protocols~\cite{EDHOCPresentation,Edhocsecuirytnalysis}. 

As a result, \gls{edhoc} can run on IoT devices needing low energy consuming protocols as well as for Vehicle-to-Vehicle (V2V) or Vehicle-to-Infrastructure (V2I) structures. In the future, \gls{edhoc} could also be used as a B5G networks standard association protocol, for example between a Base Station and a User Equipment, to guarantee secure communications~\cite{EDHOCPresentation}.  The basic version of the protocol is based on three mandatory messages as well as an optional fourth. 

\begin{figure}[!b]
  \begin{center}
    \includegraphics[width=\linewidth]{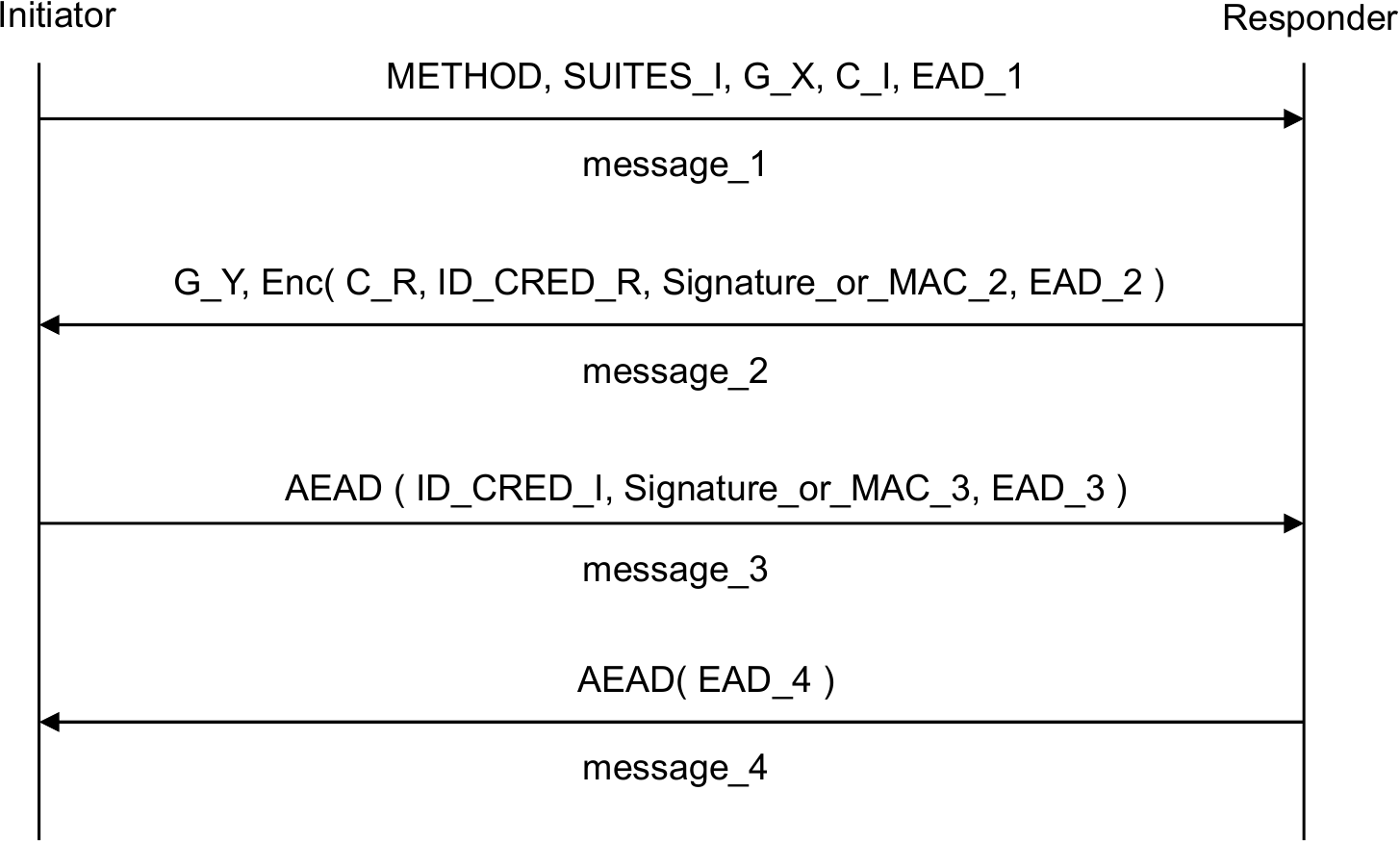}
  \end{center}
  \caption{Extract of an \gls{edhoc} message exchange and their contents, as defined in~\cite{EdhocRFC9528}.}
  \label{fig:edhoc}
\end{figure}

Figure~\ref{fig:edhoc} shows a sketch of the \gls{edhoc} message flow and their content. In the \emph{first message}, from the Initiator (I) to the responder (R), I indicates the available methods and cryptographic suites as well as its Elliptic Curve Diffie-Hellman (ECDH) ephemeral public key (G\_X) and a connection identifier (C\_I) used to associate the public key to a specific connection, in the case of multiple connections at the same time on the same machine. In this first message, as well as in future messages, the EAD\_n field is always used for External Authorization Data, for applications to directly put their security specifications into the \gls{edhoc} protocol. This field has a great impact on the size of the message and can be ignored by the Responder, respectively the Initiator, in the case where it is not specified as critical~\cite{EdhocRFC9528}.

In the \emph{second message}, R replies with its own public key (G\_Y) and connection identifier (C\_R) as well as with the ID\_CRED\_R and Signature\_and\_MAC fields used for authentication. The two last fields verify proof of possession of the private key in the different authentications methods, specified later. The second message is also composed of the next EAD field.

The \emph{third message} should be encrypted as per Authenticated Encryption with Associated Data, meaning that it will encrypt its content with the key of the \gls{edhoc} protocol as well as with the data from the optional previous EAD fields. With this third message, authentication of the Initiator is performed. The message contains the ID\_CRED\_R as well as the Signature\_and\_MAC data of the Responder.  The \emph{fourth optional message} is used to strengthen authentication, especially in the case where static Diffie Hellman keys are used. 

\section{Authentication and Secrecy Properties}

\noindent To authenticate the devices participating in the exchange, \gls{edhoc} has four original options, as well as a fifth one under review, presented in Table~\ref{tab:edhocauth} (cf.\cite{EDHOCPresentation,EdhocRFC9528,EDHOCPskDraft} and citations thereof). These five authentication methods are relevant in the context of MitM attacks, since they may decide whether the attack would succeed or fail.

The basic version of the \gls{edhoc} protocol mentions weak resistance with respect to \gls{pcs} properties. \gls{pcs}, often referred as well as backwards secrecy, refers to the capacity of a cryptographic protocol to \emph{heal} back after corruption or compromise of long-term  authentication credentials. It can also be defined as the capacity of a cryptographic protocol to bounce back after a given, finite interval, even when corruption of credentials happens, hence affecting the integrity of the communication channel between users~\cite{PCSDef}. 

Yet, a distinction is usually made between weak \gls{pcs} and perfect \gls{pcs}. The difference is made in the type of keys the adversary has access to. Weak \gls{pcs} guarantees that once an adversary loses access to a party’s secrets for one specific session, it can no longer decrypt future sessions. Perfect \gls{pcs}, on the other hand, would imply that even if long-term keys of a user have been compromised, the protocol is able to \emph{heal} and restore confidentiality, i.e., making sure that no future sessions can be decrypted. Achieving perfect \gls{pcs} is extremely demanding, since the idea persists that if an attacker has access to long-term keys of one user, it can then compute every same operation than the user to compute future session keys for each session. In practice, almost no lightweight protocol fully resists such attacks~\cite{NoPerfectPCS}. \gls{edhoc}, like many other modern key exchange mechanisms, provides forward secrecy but does not achieve perfect \gls{pcs}, meaning that a compromise at a given moment of the long-term credentials has long-lasting consequences on the security of the channel.

\section{Lawful Interception of Traffic} 
\label{subsec:LI}

\noindent Results in~\cite{Lakers} focus on \gls{edhoc} compliance with respect to \gls{li}, i.e., technical  implementation of communication channels surveillance (e.g., Patriot Act in the US, granting federal authorities access to digital data owned by companies or private users, without their consent or notification). \gls{li} raises many questions, both societal and technical. If an authority has access to supposedly encrypted data, is this data truly protected, and could other actors gain access to it as well? More broadly, if an authority can decrypt user data at will, can that data still be considered secure? Ultimately, this is a question of trust: depending on the state in which a user lives, to what extent can authorities be considered entirely trustworthy?  

Current research on \gls{li} focuses on enabling authorized authorities to access compromising data without entirely breaking end-to-end encryption. At the same time, the goal is to avoid drifting into mass surveillance, which can be defined as a situation where anyone can be subject to arbitrary interference in their privacy. Correct use of \gls{li} should be limited to non-arbitrary cases, and ideally to cases where the user has explicitly given consent.  

In modern proposals for \gls{li}, the relevant problem is how to make the session key available to a third party, the lawful interceptor, without violating the principle of end-to-end encryption. The most common approach today involves key escrow~\cite{KeyEscrow}, but this technique poses a fundamental problem: it requires the third party to have direct access to the encryption key at any time, which is essentially incompatible with the idea of end-to-end protection. Work in~\cite{Lakers} instead presents alternative methods, for example using three different keys: two from the users and one from the authorities. In this case, the authority must request the users’ cooperation to obtain their keys before decryption can occur.  


\section{LI-Compliance for \gls{edhoc}}

\noindent While the protocol is still in the standardization process \cite{EdhocRFC9528}, the LAKE working group of the IETF has already explored how to implement \gls{li}-compliance in \gls{edhoc} \cite{Lakers}. This is a particularly interesting use case, since it attempts to incorporate \gls{li} while still avoiding mass surveillance scenarios. In their proposal, the Initiator and Responder perform a standard \gls{edhoc} exchange without modification, while at the same time each protocol message is sent to a proxy and encrypted with the public keys of the participants and the authorities. To perform interception and recover the shared secret of \gls{edhoc}, all parties --including the Initiator, the Responder, and the authorized authorities -- must contribute their respective secrets derived from their private keys.  

This approach has several advantages. The endpoints run a standard \gls{edhoc} exchange, preserving compatibility and avoiding disruption of existing implementations. Interception is fine-grained, limited to the targeted session rather than enabling bulk surveillance. The requirement that all parties provide their share of the secret ensures that no single actor can unilaterally recover the key, mitigating the risk of indiscriminate monitoring. The scheme also maintains \gls{edhoc}’s identity-protection guarantees, making it impossible to falsely attribute participation in a session to an innocent party.  

Nevertheless, important limitations remain. Additional cryptographic operations, such as secret encapsulation and proofs of knowledge, add computational overhead, which may be prohibitive for constrained IoT devices. Furthermore, interception depends on the cooperation of all designated authorities: if even one refuses or is unavailable, interception cannot occur. The reliance on proxies and supporting infrastructure also creates residual trust issues, since any compromise of these entities could endanger user privacy. Finally, as the proposal is still under discussion, questions remain about its scalability and feasibility for large-scale deployment. More broadly, while the design reduces the risk of blanket surveillance, it nevertheless introduces a form of backdoor. 

\section{Implementation of the Attack}

\subsection{Assumptions}
\label{sec:assu}

\noindent Our attack scenario assumes compromised long-term authentication credentials of the parties or key escrow techniques (e.g., known pre-shared cryptographic keys held in trust, due to legally mandated situations). The same attack assumption would break any other centralized PKI-based authentication protocol. 

\subsection{Attack Scenario}

\noindent We assume a malicious device able to act simultaneously as a malicious rely between two nodes. The goal would be to make an endpoint believe that it is communicating with a legitimate node  and a legitimate node believe that it is communicating with an endpoint, similar to how a MitM attack between an end point and a base station would work. We can also consider a setting in which the malicious device has been configured to facilitate \gls{li}, with authorities treated as potential adversaries. The aim is not to defeat end-to-end cryptography, but to demonstrate how programmable propagation can make interception easier, more reliable, and less visible. In the end, we assume three endpoints, two acting as legitimate nodes and one acting as the malicious interception point. 

Our scenario gets inspiration on the use of malicious metasurfaces conducting the attack presented in~\cite{shaikhanov2022metasurface}, referred to as Metasurface-in-the-Middle (MSitM) attack. The technical part of the attack is inspired from the CoopeRIS simulation framework~\cite{cooperis}, which models metasurface-assisted vehicular communications over  OMNeT++~\cite{varga2010omnet++}. More precisely, CoopeRIS provides the specific case of \gls{ris}, as an emerging solution in B5G scenarios, and composed of electromagnetic elements capable of altering the phase of incident of radio waves, allowing them to actively shape and redirect the wireless environment in real time~\cite{RISSoftwareEnvironment,RIScontrolint,RISPotential}. Building our scenario over CoopeRIS, several attack strategies were designed and implemented, including malicious traffic redirection, side-lobe eavesdropping, \gls{ris}-based MitM attacks, and \gls{ris}-assisted lawful interception. While the complexity of the framework limited some aspects of physical-layer modeling, these simulations provided valuable insights into how \gls{ris} may reshape both attack surfaces and defensive strategies in future wireless environments.

\begin{figure}[!h]
  \begin{center}
    \includegraphics[width=\linewidth]{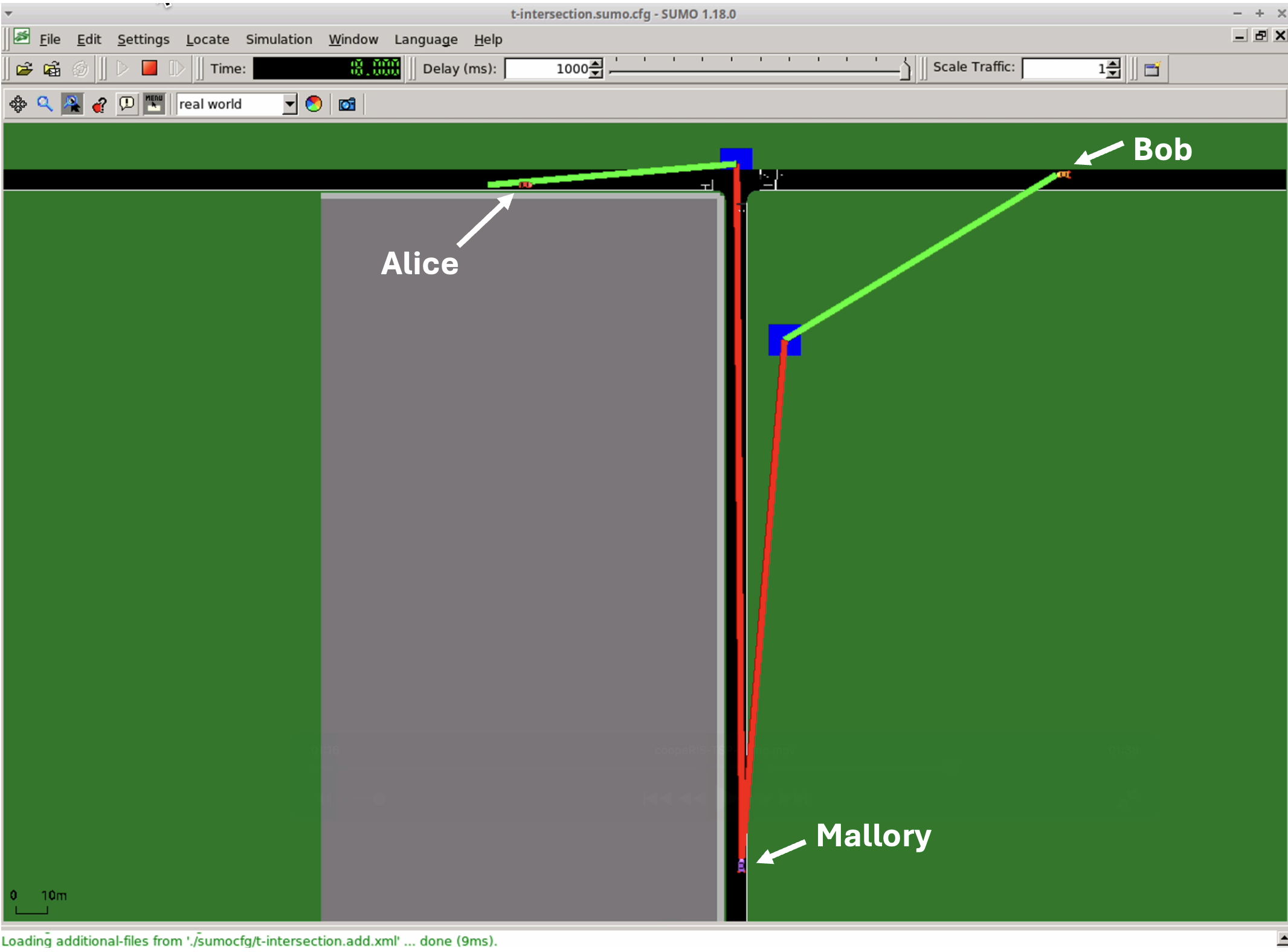}  
   \end{center}
  \caption{\gls{ris}-assisted lawful interception attack scenario  (cf. companion github repository~\cite{MitMEdhocGit} for further details and videocaptures associated to the attack).}
  \label{fig:final-scenario}
\end{figure}

\begin{figure}[!h]
  \begin{center}
    \includegraphics[width=\linewidth]{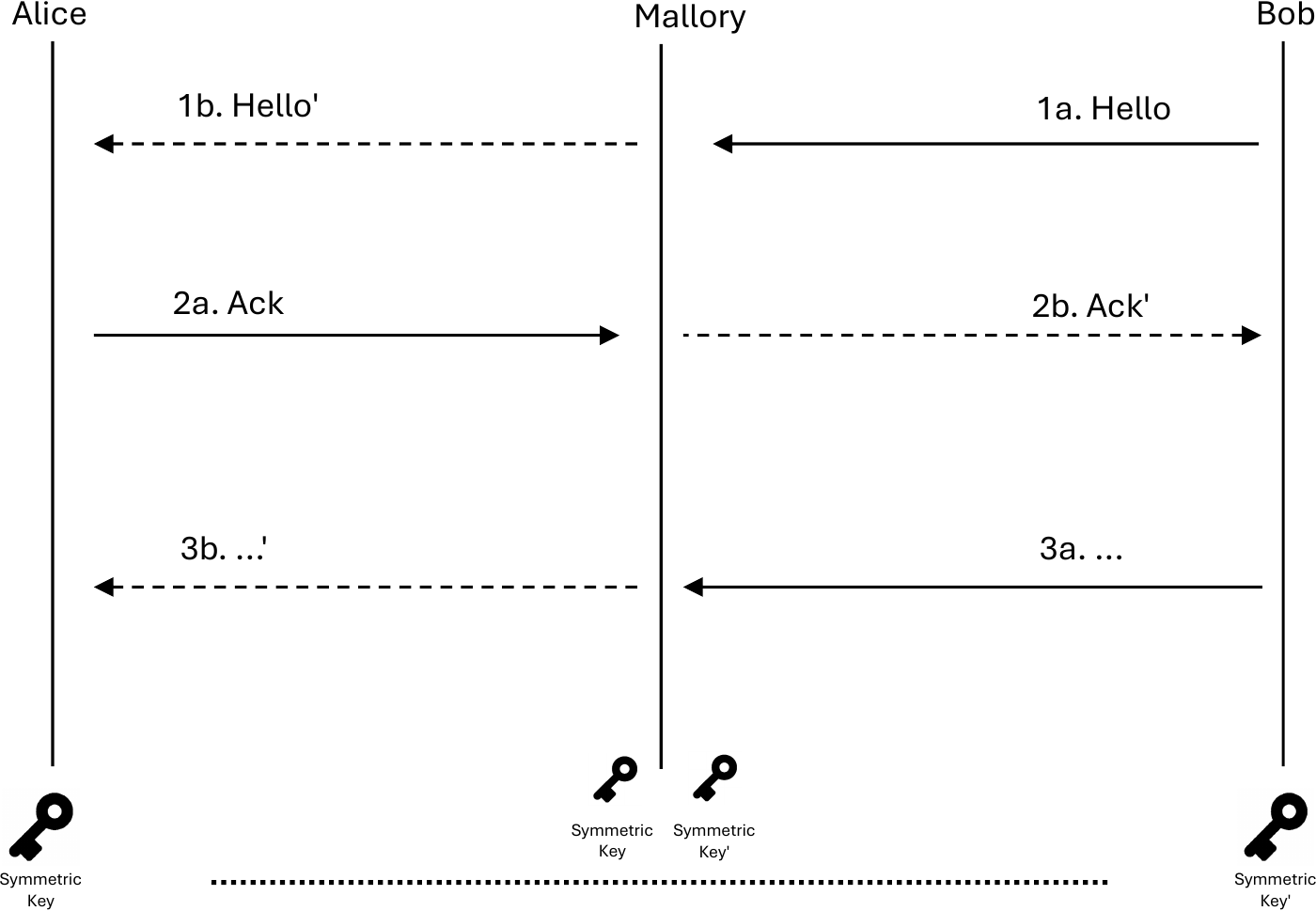}
  \end{center}
  \caption{Sample representation of the attack w.r.t. the intercepted and modified message exchange flows.}
  \label{fig:edhocMSitM}
\end{figure}

One metasurface is configured to illuminate the proxy consistently, while the other serves the legitimate receiver. This way, communications are both delivered and mirrored without degrading user-perceived quality. In such a deployment, users may remain unaware that traffic is being duplicated, because the \emph{extra} path is realized in the physical layer rather than by inserting an active relay in the logical network path. This configuration gives a visual representation and simulation of the scenario, to get a grasp of how it could work. Although a \gls{ris} cannot decrypt robust end-to-end sessions such as those established with modern \gls{edhoc} authentication, it could serve as a way to facilitate global capture of data for interception, for example if there is only one proxy situated in a very large city for the sake of being more discrete, multiple \gls{ris} having a benign purpose initially could be used simultaneously to send all of their traffic towards the proxy as well. Figures~\ref{fig:final-scenario} and \ref{fig:edhocMSitM} depicts the general idea of our \gls{ris}-assisted lawful interception attack.

The implementation associated to Figures~\ref{fig:final-scenario} and \ref{fig:edhocMSitM} reuse existing code by Krontiris in~\cite{krontirisEDHOC}, which provides a solid base of the \gls{edhoc} protocol in C language, which we modified under the context of the CoopeRIS framework~\cite{cooperis} to simulate the attack PoC. The code was modified to ensure weak authentication between nodes, as defined in our assumptions (cf. Subsection~\ref{sec:assu}). The full code, together with some videocaptures showing the most results of the attack, can be found at our companion github repository\footnote{Available online, cf.~\cite{MitMEdhocGit}} to foster further research on the topic.

\section{Perspectives for future work}

\noindent The issues reported in this document may raise the following research questions: how can endpoints or higher network layers reliably identify whether traffic is being redirected through malicious entities? One possible approach might involve the use of anomaly detection at the physical layer, e.g., by comparing expected propagation models with measured signal statistics.  Another potential solution is the use of encryption over anamorphic channels~\cite{persiano2022anamorphic,banfi2024anamorphic,Cinal2025anamorphic}, which could allow ciphertexts being decrypted into different messages with some sort of out-of-band verification (i.e., by concealing messages from the real content to the eyes of the surveillance framework). Some other practical solutions could rely on the use of decentralized \gls{pki} schemes, ensuring the use of different authorization platforms.\\

\noindent  \textbf{Acknowledgments ---} The authors would like to acknowledge fruitful discussions on the topics of this work with the following people (in alphabetical order): M. Barbeau, L. De Cicco, P. Leleux, E. Lopez-Perez, C. Onete, A. Pierron, J. Rubio-Hernan, M. Segata, M. Vučinić. The  work has been partially supported by the French National Research Agency under the France 2030 label (NF-HiSec ANR-22-PEFT-0009).

\bibliographystyle{unsrt}

\end{document}